\documentclass[twocolumn,amsmath,amssymb]{revtex4}
\usepackage{graphicx}% Include figure files
\usepackage{dcolumn}% Align table columns on decimal point
\usepackage{bm}% bold math
\begin{document}

\title{Superluminal Tachyonlike Excitations of Dirac Fermions in
a Topological Insulator Junction
}
\author{Vadim M. Apalkov}
\affiliation{Department of Physics and Astronomy, Georgia State University,
Atlanta, Georgia 30303, USA}
\author{Tapash Chakraborty$^\ddag$}
\affiliation{Department of Physics and Astronomy,
University of Manitoba, Winnipeg, Canada R3T 2N2}

\date{\today}
\begin{abstract}
We have considered a system of two topological insulators and have 
determined the properties of the surface states at the junction.
Here we report that these states, under certain conditions exhibit
superluminous (tachyonic) dispersion of the Dirac fermions. Although
superluminal excitations are known to exist in optical systems, this 
is the first demonstration of possible tachyonic excitations in a purely
electronic system. The first ever signature of tachyons could therefore
be found experimentally in a topological insulator junction.
\end{abstract}
%\pacs{}
\maketitle

Topological insulators (TIs), a new class of materials rich with new
concepts and promises, have attracted considerable attention in the
condensed matter physics community \cite{topo_review}. In the bulk, the 
system is electrically insulating, driven by the strong spin-orbit coupling 
present in the system. The three-dimensional (3D) TIs host a single 
Dirac cone in the surface states that was confirmed experimentally 
by angle-resolved photoemission spectroscopy. Dirac fermions are also 
present in graphene \cite{abergeletal}. However, in the TIs, unlike in 
graphene, there exists only odd number of non-degenerate Dirac cones 
with spin-momentum locking that results in helical Dirac fermions 
\cite{helical} without spin degeneracy. This spin chirality of Dirac
fermions prevents them from backscattering and localization \cite{roushan}. 
That makes those systems ideal for spintronics applications or for
quantum computing \cite{topo_review}. Until now, most of the attentions 
have been heaped on the surface states of a single TI. Here we show that, 
the junction surface states of two TIs, in certain situations, exhibit 
superluminal (tachyonic) dispersion of Dirac fermions. Tachyons have 
eluded detection until now, despite diligent efforts by the particle 
physicists worldwide. However, as we have demonstrated below, it could 
perhaps be found in the present solid state system.

We consider a junction between two topological insulators. The
junction surface is described by $z=0$, and we assume
that the system of TIs is isotropic in $x$ and $y$ directions.
Therefore, the surface states are characterized by the $x$ and $y$
components of the wave vector, $k^{}_x$ and $k^{}_y$, and the surface
wave functions depend on the $z$-coordinate and decay in both directions,
i.e., the positive and negative directions of the $z$ axis [see inset
in Fig.~1(b)]. We label the topological insulator at $z <0$ as TI-1, 
and the topological insulator at $z >0$ as TI-2. We have found that
for a general variation of parameters of TI-2, the junction surface
states exhibit one branch with unique {\it tachyonic} dispersion
relation. Although, they are not yet found experimentally, these
``\"Uberlichtgeschwindigkeitteilchen" (faster than light particles)
discussed by Sommerfeld \cite{sommer} in 1905, and many others
\cite{tachyonics} since then, have always been vigorously
pursued (e.g., in the case of the neutrinos \cite{chodos,ehrlich}) by the
particle physics community for many decades \cite{foundation}.

We assume that the electronic states of both TIs are described by
the same type of low-energy effective 3D Hamiltonian
\cite{liu_2010,zhang_2009}, which has the $4\times 4$ matrix form
and can be expressed as
\begin{equation}
{\cal H}^{}_{\rm TI}=\varepsilon(\vec{k})+ \left(\begin{array}{cc}
 M(\vec{k})\sigma^{}_z- {\rm i} A^{}_1 \sigma^{}_x \partial_z & A^{}_2
k^{}_{-}\sigma^{}_x \\
A^{}_2 k^{}_{+}\sigma^{}_x & M(\vec{k})\sigma^{}_z + {\rm i} A^{}_1
\sigma^{}_x \partial_z
\end{array}\right),
\label{HTI}
\end{equation}
where $\sigma^{}_i$ ($i=x,y,z$) are the Pauli matrices, $\partial_z
= \partial/\partial z$, $\vec{k} = (k_x, k_y)$ is a two-dimensional (2D)
wave vector, $k^{}_{\pm} = k^{}_x \pm {\rm i} k^{}_y$, and
\begin{eqnarray}
& & \varepsilon (\vec{k})=C^{}_1 - D^{}_1 \partial_z^2 + D^{}_2 (k_x^2
+ k_y^2), \\
& & M(\vec{k})= M^{}_0 + B^{}_1 \partial_z^2 - B^{}_2 (k_x^2 + k_y^2).
\end{eqnarray}
For a topological insulator of the type $\mbox{Bi}^{}_2\mbox{Se}^{}_3$,
the four-component wave functions, $\Psi$, corresponding to the matrix
Hamiltonian (\ref{HTI}) determine the amplitudes of the wave functions
at the positions of Bi and Se atoms: $(\mbox{Bi}^{}_{\uparrow},
\mbox{Se}^{}_{\uparrow}, \mbox{Bi}^{}_{\downarrow},
\mbox{Se}^{}_{\downarrow})$, where the arrows indicate the direction
of the electron spin. In the case of $\mbox{Bi}^{}_2\mbox{Se}^{}_3$ TI,
the
constants in the Hamiltonian (\ref{HTI})  are \cite{zhang_2009},
$A^{}_1 = 2.2$ eV$\cdot$\AA,  $A^{}_2 = 4.1$ eV$\cdot$\AA,
$B^{}_1 = 10$ eV$\cdot$\AA$^2$, $B^{}_2 = 56.6$ eV$\cdot$\AA$^2$,
$C^{}_1 = -0.0068$ eV,
$D^{}_1 = 1.3 $ eV$\cdot$\AA$^2$,
$D^{}_2 = 19.6 $ eV$\cdot$\AA$^2$, and
$M^{}_0 = 0.28$ eV.

The unique properties of the bulk Hamiltonian (\ref{HTI}) is that, for
a single TI, it can produce surface states with massless relativistic
dispersion relation, $E \approx v^{}_{\rm F} k$, where $v^{}_{\rm
F} = A^{}_2 \sqrt{1 - (D_1/B_1)^2}$ \cite{shan_2010} is the Fermi velocity. 
In the case of two TIs, at the junction we expect a coupling between 
two surface states belonging to different TIs. Within a simple model which
includes phenomenological coupling between massless relativistic
states of the two TIs, it was shown earlier that the properties of
the junction surface states strongly depend on the relative sign of
the Fermi velocities of the two TIs, i.e., on the relative sign of $A^{}_2$
for TI-1 and TI-2 \cite{takahashi}. Here we show that for a realistic
3D model [Eq.~(\ref{HTI})] of the TI, one can observe new and unique 
features in the dispersion relation of the junction states. 
 
In what follows, we study the junction surface states within the
realistic 3D model of the TI. For two TIs, we assume that both TIs are
described by the Hamiltonian of the same type (\ref{HTI}) but
with different constants. To distinguish the constants corresponding
to different TIs, we introduce superscripts $(1)$ and $(2)$ for TI-1
and TI-2, respectively. Following the general procedure of constructing 
the surface states of a TI \cite{shan_2010,zhou_2008}, we first determine 
for each TI the general bulk solution of the Schr\"{o}dinger equation 
of the form $\Psi\propto {\rm e}^{\lambda^{(m)} z}\, {\rm e}^{{\rm
i}\vec{k}\vec{\rho}}$, where $m=1$ and 2 for TI-1 and TI-2, respectively. 
Substituting this form of solution in the Schr\"{o}dinger equation, 
${\cal{H}}_{\rm TI} (\vec{k},\partial_z)\Psi = E\Psi$, we obtain a secular 
equation, $\det\left[{\cal{H}}^{(m)}_{\rm TI} (\vec{k},\lambda^{(m)})
-E\right]=0$, for each TI ($m=1,2$). For each energy $E$, this equation defines
four values of $\lambda^{(m)}_{\alpha }(k, E)$, $\alpha=1,\ldots,4$.
Each $\lambda^{(m)}_{\alpha }(k, E)$ is doubly degenerate, which
finally generates eight wave functions for each TI, $m=1$ and 2,
$\Psi^{(m)}_{s,\alpha} (k,E)\, {\rm e}^{\lambda^{(m)}_{\alpha} z}
{\rm e}^{{\rm i} \vec{k}\vec{\rho}}$ (see Ref.~\cite{shan_2010,zhou_2008}),
where $s=1,2$, $\alpha=1,\ldots,4$, and $\Psi^{(m)}_{s,\alpha} (k,E)$
is a four-component wave function.

Our goal here is to determine the surface states localized at the
junction
of the two TIs. Therefore, out of the four values of $\lambda^{(m)}$
(for each TI), we choose only two values: for TI-1 ($z < 0$) with
$\mbox{Re}
\lambda^{(m)}_{\alpha} > 0 $ and for TI-2 ($z > 0$) with $\mbox{Re}
\lambda^{(m)}_{\alpha} <0$. After selection of these $\lambda^{(m)}_{\alpha}$
($\alpha = 1, 2$), our junction surface states then take the form
\begin{equation}
\Phi(\vec{\rho},z) = \left\{
\begin{array}{l}
\sum\limits_{{s=1,2}\atop{\alpha =1,2}} C^{(1)}_{s,\alpha}
\Psi^{(1)}_{s,\alpha}\, {\rm e}^{\lambda^{(1)}_{\alpha} z} {\rm e}^{{\rm
i}
\vec{k}\vec{\rho }} , z <0  \\
\sum\limits_{{s=1,2}\atop{\alpha=1,2}} C^{(2)}_{s,\alpha}
\Psi^{(2)}_{s,\alpha} {\rm e}^{\lambda ^{(2)}_{\alpha}z}
\,{\rm e}^{{\rm i} \vec{k}\vec{\rho}}, z >0.
\end{array}
\right.
\end{equation}
This type of wave function determines the localized junction
surface states. The energy of the junction state is found from
the condition of continuity of the wave function, $\Phi(z)$, and
the corresponding current, $[\delta {\cal H}^{(m)}_{\rm TI}/\delta
k^{}_z] \Phi(z),\ (k^{}_z= {\rm i}\partial_z$) at the junction between
the two TIs. Solution of this continuity equations determines the
dispersion relation, $E(k)$, of the junction surface states.

\begin{figure}
\begin{center}\includegraphics[width=7cm]{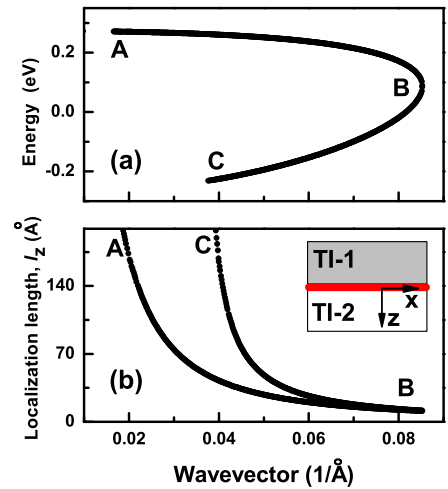}
\end{center}
\caption{\label{Fig2}
(a) Dispersion relation of the junction states. The parameters of the
TI-2 are the same as the parameters of TI-1 except for $A^{(2)}_1 = 
2.2$ eV$\cdot$\AA. Near the point B the dispersion relation shows
the tachyonic behavior with superluminal group velocity.
(b) The localization length of the junction states shown in panel (a).
Near the tachyonic point (point B) the junction states are strongly
localized with localization length $\approx 10$ \AA. At the edge of the
tachyonic branch (near points A and C) the states become delocalized.
Inset: Schematic illustration of the junction (red line) between two TIs.
The coordinate system is introduced as follows: $z$-axis is perpendicular 
to the junction surface, and $x$ and $y$ are in the plane of the junction.
}
\end{figure}

We keep the parameters of TI-1 fixed as for $\mbox{Bi}^{}_2 \mbox{Se}^{}_3$ 
(as given above) and vary the parameters of TI-2. Our results indicate that
for general variation of the parameters of TI-2, the junction surface
states show one branch with unique dispersion relation that resembles
the dispersion of the tachyons \cite{chiao96}. In Fig.~1 we show the
dispersion relation for the junction state when only one parameter,
$A_1$, of TI-2 is different from TI-1. In this case there is only one
type (tachyonic) of dispersion. The group velocity,
$v^{}_g = \partial E(k)/\partial k$, corresponding to the tachyonic
dispersion becomes infinitely large at $k \approx k_0$ [point B in
Fig.~1(a)]. Similar superluminal dispersion was known to be present
(theoretically) for propagation of the optical pulses through inverted 
two-level system of atoms \cite{chiao96} and in metamaterial photonic 
crystals \cite{chen11} with folded bands. Superluminal propagation
can also be observed for propagation of light pulses through a
media with anomalous dispersion relation \cite{milonni02}. In all
these cases, the tachyonic dispersion is achieved for propagation of
light pulses through a specially designed medium. In contrast, our system
consists of just electronic degrees of freedom and propagation of
electronic excitations exhibits tachyonic dispersion. If confirmed
experimentally, this would be the first example of superluminal
tachyonic dispersion in an actual semiconductor (electronic) system.

In Fig.~1(b) we show the localization length in the $z$-direction of the 
junction states of the tachyonic branch. The localization length is 
defined as $l^{}_z=\mbox{min}\left\{ 1/|\mbox{Re}\lambda^{(m)}_{s,\alpha} 
|\right\}$. Clearly, the maximum localization of the surface states
occur near the tachyonic point B with localization length $\approx 10 $ \AA. 
These strongly localized states are less sensitive to the bulk disorder 
of the TIs, which can help with the experimental observation of the tachyonic 
states. Away from the tachyonic point the junction states become weakly 
localized and finally, near the ends of the tachyonic branch (points A 
and C) the junction states become delocalized. The actual distribution 
of electronic density for the two (up and down) spin components, is shown 
in Fig.~2 for one of the junction states near point B [see Fig.~1(a)]. 
Both spin components are occupied with final non-zero spin-polazition of 
the junction tachyonic states. Exactly at the tachyonic point the junction
state is spin unpolarized.

\begin{figure}\begin{center}\includegraphics[width=7cm]{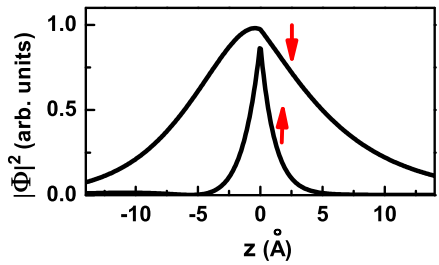}
\end{center}\caption{\label{Fig3}
The electron density along the $z$ direction for the tachyonic branch 
shown in Fig.~1(a). The density is shown for the junction surface state
at $k=0.084$ \AA$^{-1}$. The arrows next to the lines are the directions 
of the electron spin. The junction between the TIs is at $z=0$.
}\end{figure}

The existence of a tachyonic branch in the dispersion relation of junction 
states means that the energy dispersion becomes a non-analytic function 
of the 2D surface momentum, i.e., the group velocity $v^{}_g = \hbar^{-1} 
\partial E(k)/\partial k$ is infinitely large. Due to the analytical dependence 
of the Hamiltonian (\ref{HTI}) on the wave vector $\vec{k}$, the non-analyticity 
in the dispersion relation is possible only if the Hamiltonian is non-hermitian. 
In our case the junction states are decaying states, i.e., $\Phi \propto 
{\rm e}^{\lambda_{\alpha} z}$, where the real part of $\lambda_{\alpha}$ 
is non-zero. For these states the Hamiltonian becomes non-hermitian and 
tachyonic branches are therefore allowed. The existence of tachyonic 
dispersion does not violate Einstein's causality principle. The reason 
is that the superluminal group velocity describes the propagation of 
not a signal but an analytical wave packet. The propagation of a singularity, 
i.e., the signal, is not described by the group velocity, so there is no 
violation of Einstein's causality relation (see Ref.~\cite{chiao96,milonni02}). 

The tachyonic excitations can be described by an effective 2D
Hamiltonian. This effective Hamiltonian should have a $2\times 2$ matrix form, 
which takes into account the electron spin degrees of freedom. In addition, the 
Hamiltonian should also support the non-analytical tachyonic dispersion
relation and should be non-hermitian. The tachyonic Hamiltonian can be 
constructed, for example, by introducing the imaginary proper mass in 
the Dirac equation \cite{jentschura12}. In our case, the tachyonic
Hamiltonian is achieved by introducing an imaginary Fermi velocity in the 
massive Dirac equation. More precisely, the effective Hamiltonian 
which describes the tachyonic junction surface states has the form 
\begin{equation}
{\cal H}^{}_{\mbox{Tach}} = \left( 
\begin{array}{cc}
\Delta_0 & {\rm i} \hbar v^{}_{\rm I} k^{}_{+}  \\
{\rm i} \hbar v^{}_{\rm I} k^{}_{-} & - \Delta^{}_0     
\end{array}
\right)  = \Delta^{}_0 \sigma^{}_z + {\rm i} \hbar v^{}_{\rm I} 
(\vec{\sigma} \vec{k}),
\label{Heff}
\end{equation} 
where $\Delta^{}_0$ is the effective mass of the tachyons, and 
${\rm i} v^{}_{\rm I}$ is the imaginary Fermi velocity. This effective 
Hamiltonian produces a tachyonic branch with dispersion relation of the form
\begin{equation}
E^{}_{\mbox{Tach}} (k) = \pm \sqrt{\Delta_0^2 - \hbar^2 v_{\rm I}^2 k^2}.
\end{equation}
Therefore, $k<k^{}_0=\Delta^{}_0/\hbar v^{}_{\rm I}$ and the group velocity at 
$k = k^{}_0$ becomes infinitely large. For the tachyonic branch shown in 
Fig.~1 the parameters of the effective Hamiltonian (\ref{Heff}) are 
$\Delta^{}_0 = 0.313 $ eV and $v^{}_{\rm I} = 5.7\times 10^5$ m/s. 

The wave function corresponding to Hamiltonian (5) with energy spectrum (6) has
the following form
\begin{equation}
\Psi^{}_{\mbox{Tach}} = \left(
\begin{array}{c}
{\rm e}^{{\rm i}\phi/2} \cos \alpha (k) \\
{\rm i}{\rm e}^{-{\rm i}\phi/2} \sin \alpha (k)
\end{array}
\right),
\end{equation}
where $k=\sqrt {k_x^2 + k_y ^2}$, $ \phi = \cos^{-1} k^{}_x /k $, and
$\alpha (k)=\sin^{-1}\sqrt{\frac{\Delta^{}_0 - E^{}_{\rm Tach} (k)}{2\Delta^{}_0}}$.
The corresponding direction, $\vec{n}$, of electron spin is characterized by
an angle $2\alpha$ relative to the $z$ axis and is given by $\vec{n}=
(n^{}_x,n^{}_y,n^{}_z)=(\sin 2\alpha \sin \phi, \sin 2\alpha 
\cos \phi, \cos 2\alpha)$. Therefore, for a given tachyonic state, the 
$z$-component of the electron spin is $\cos2\alpha(k)$, i.e., the state is 
spin polarized. At $k=k^{}_0$, i.e., at a singular point of the tachyonic branch, 
the angle $\alpha = 45^0$ and the electron state is spin unpolarized. This behavior 
is consistent with the exact distribution of the electron density shown in Fig.~2, 
which illustrates finite spin polarization of the tachyonic state away from 
the singular point B ($k=k^{}_0$).

The results shown in Fig.~1 illustrate the existence of tachyonic branches 
under variation of just one parameter, $A^{}_1$, of TI-2. By varying the 
other parameters, we can introduce additional junction states with more 
than one tachyonic branch. If the signs of the constants $A^{}_2$ for 
two TIs, i.e., the signs of the Fermi velocities for two isolated TIs
are the same, then usually we observe two tachyonic branches as shown 
in Fig.~3(a). If the signs of $A^{}_2$ are opposite then the junction 
states usually have one or two tachyonic branches and one massless 
relativistic Dirac branch [see Fig.~3(b)]. This massless relativistic 
branch was predicted earlier within a 2D model Hamiltonian of the 
junction states \cite{takahashi}. 

\begin{figure}
\begin{center}\includegraphics[width=8cm]{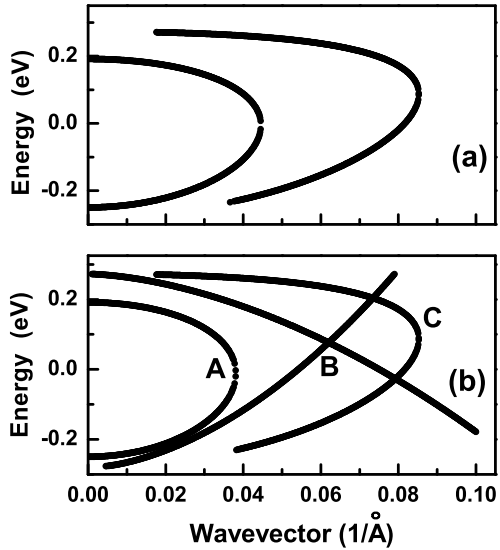}
\end{center}
\caption{\label{Fig4}
Energy dispersion curves for the junction states. The parameters of the 
TI-2 are the same as the parameters of TI-1 except 
(a) $A^{(2)}_1 = 3.0$ eV$\cdot$\AA and $A^{}_2 = 5.1$ eV$\cdot$\AA;
(b) $A^{(2)}_1 = 3.0$ eV$\cdot$\AA and $A^{}_2 = -4.1$ eV$\cdot$\AA. 
Branches A and C correspond to the tachyonic dispersion, while the 
branch B describes massless relativistic Dirac fermions. 
}
\end{figure}

As the topological insulators with different Fermi velocities are
already available in the laboratories, perhaps we could propose
ways to detect the elusive tachyons in these systems. One possible 
physical manifestation of tachyonic dispersion relation in our
system would be the display of a singularity in the time resolved 
measurements of 2D ballistic electron transport \cite{shaner04} along
the junction between the TIs. Under the ballistic condition the electron 
wave packet propagates with a group velocity, which is determined by the
corresponding dispersion relation. Within the effective model of tachyonic 
branch this group velocity is
$$v^{}_g = \hbar^{-1} \partial E_{\rm Tach}(k)/\partial k = \pm\frac{\hbar 
v_{\rm I}^2 k}{\sqrt{\Delta_0^2 - \hbar^2 v_{\rm I}^2 k^2}},$$
where the positive (negative) group velocity corresponds to electron
(hole) excitation.
Therefore the ballistic transport time through a finite distance, $L^{}_b$, is 
$t^{}_b = L^{}_b/v^{}_g \propto \sqrt{k_0 ^2 - k^2}$. At a point $k = k^{}_0$, 
i.e., where the group velocity is infinitely large (point B in Fig.~1) 
the transport time becomes very small. At this point the transport time
as a function of the energy would exhibit a cusp-like singularity. That means,
at that energy the time of electron transport will have a sharp minimum. The actual
singularity will be smeared due to a finite width in $k$-space of the electron wave packet.

\begin{figure}
\begin{center}\includegraphics[width=7.5cm]{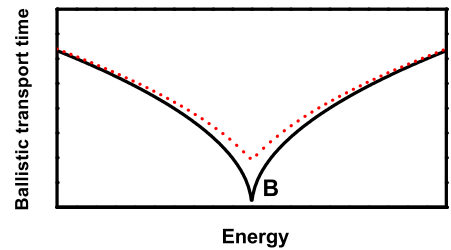}
\end{center}
\caption{\label{Fig5}
Schematic illustration of the ballistic propagation of tachyonic 
excitations (electrons with positive group velocity and holes with negative
group velocity) through the junction layer between two TIs. 
At the point of diverging group velocity, the ballistic travel time as a 
function of electron energy shows a cusp-like singularity (solid line).
The finite width of the electron wave packet results in smearing of 
this singularity (dotted line). Point B corresponds to the singular
point in Fig.~1.
}
\end{figure}

In conclusion, we have shown here that for a general set of parameters
for the topological insulator Hamiltonian, the surface states at the 
junction between two TIs have at least one unique tachyonic branch, 
which describes the propagation of Direc fermion excitations with superluminal
group velocity. Although the excitations propagating with superluminal velocity 
are known (theoretically) to exist in optical systems, we show here that 
such excitations can actually be realized in purely electronic systems. 
In these systems, the tachyonic excitations are well localized at the 
junction between the TIs and are susceptible to direct experimental observation. 
It would indeed be a remarkable feat for condensed matter and the materials 
sciences if the first ever signature of elusive tachyons is actually detected 
experimentally in a topological insulator junction.

The work has been supported by the Canada Research Chairs Program of the 
Government of Canada.

\end{document}